\documentclass[apjl,a4paper,12pt,useAMS]{emulateapj}
\usepackage{amsmath}
\usepackage{cases}
\usepackage{amssymb}
\usepackage{graphicx}
\usepackage{natbib}

\begin{document}

\title{X-ray transients from the accretion-induced collapse of white dwarfs}
\author{Yun-Wei~Yu$^{1,2}$, Aming Chen$^{1,2}$, and Xiang-Dong Li$^{3,4}$}

\altaffiltext{1}{Institute of Astrophysics, Central China Normal
University, Wuhan 430079, China, {yuyw@mail.ccnu.edu.cn}}
\altaffiltext{2}{Key Laboratory of Quark and Lepton Physics (Central
China Normal University), Ministry of Education, Wuhan 430079,
China}
\altaffiltext{3}{School of Astronomy and Space Science,
Nanjing University, Nanjing 210093, China}
\altaffiltext{4}{Key
Laboratory of Modern Astronomy and Astrophysics (Nanjing
University), Ministry of Education, China}

\begin{abstract}
The accretion-induced collapse (AIC) of a white dwarf in a binary with a nondegenerate companion can sometimes lead to the formation of a rapidly rotating and highly magnetized neutron star (NS). The spin-down of this NS can drive a powerful pulsar wind (PW) and bring out some detectable multi-wavelength emissions. On the one hand, the PW can evaporate the companion in a few days to form a torus surrounding the NS. Then, due to the blockage of the PW by the torus, a reverse shock can be formed in the wind to generate intense hard X-rays. This emission component disappears in a few weeks' time, after the torus is broken down at its inner boundary and scoured into a very thin disk. On the other hand, the interaction between the PW with an AIC ejecta can lead to a termination shock of the wind, which can produce a long-lasting soft X-ray emission component. In any case, the high-energy emissions from deep inside the system can be detected only after the AIC ejecta becomes transparent for X-rays. Meanwhile, by absorbing the X-rays, the AIC ejecta can be heated effectively and generate a fast-evolving and luminous ultraviolet (UV)/optical transient. Therefore, the predicted hard and soft X-ray emissions, associated by an UV/optical transient, provide a clear observational signature for identifying AIC events in current and future observations (e.g., AT 2018cow).
\end{abstract}

\section{Introduction}
A white dwarf (WD) in a binary with a main-sequence star, a giant, or a helium star can increase its mass toward the Chandrasekhar limit by accreting material from the companion. Therefore, such accreting WDs are widely regarded as the progenitors of Type Ia
supernovae (SNe Ia; Whelan \& Iben 1973; Nomoto 1982; Hillebrandt \& Niemeyer 2000), though SNe Ia can also originate from mergers of double WDs (Tutukov \& Yungelson, 1981; Iben \& Tutukov, 1984; Webbink 1984;
Han 1998). Nevertheless, detonation is not always inevitable for a WD close to the Chandrasekhar limit. Sometimes, if the electron captures of neon and
magnesium in the WD core take place more quickly than the
nuclear burning, the WD collapses into a neutron star (NS) (Canal \& Schatzman 1976; Miyaji et al.
1980; Canal et al. 1990; Nomoto \& Kondo 1991; Wang 2018). Such accretion-induced collapses (AICs) of WDs provide a possible route for NS formation and, in particular, provide a competitive explanation for the origin of the millisecond pulsars in low-mass X-ray binaries (Tauris et al. 2013; Liu \& Li 2017). Despite of its theoretical importance, the predicted AIC phenomenon has never been confirmed by observations. Nevertheless, in principle, the fact that some AIC candidates might have existed in the results of current transient surveys
cannot be ruled out.

It is therefore necessary to uncover some observational features of AIC events. Accompanying the WD collapse and NS formation, a small mass of $\sim10^{-3}-10^{-2}M_{\odot}$ can be ejected by a core bounce on the proto-NS (Woosley \& Baron
1992; Dessart et al. 2006) and, maybe more importantly, by a wind from the accretion disk if it rotates quickly enough (Metzger et al. 2009). Then, the collision of this AIC ejecta with the companion star can first generate an X-ray flash lasting several tens of seconds to a few hours (Piro \& Thompson 2014)\footnote{Due to the collision, a small amount of material can be stripped from the companion star, which is somewhat heated by a shock. Because of the high density of the companion, the shock velocity would not be much higher than the escaping velocity (Li \& Yu 2016) and thus the heating effect may not be as effective as considered in Piro \& Thompson (2014).}. Subsequently, on a timescale of a few days, a supernova-like transient can arise from the thermal emission of the hot ejecta heated by the radioactive decays of $^{56}$Ni (Metzger et al. 2009; Darbha et al. 2010). However, being limited by the low mass of the ejecta, the luminosity of this radioactivity-powered AIC optical emission cannot be much higher than $\sim10^{41}\rm erg~s^{-1}$, even though the synthesization of $^{56}$Ni in the ejecta is very efficient.

\begin{figure*}
\begin{center}
\includegraphics[scale=0.5]{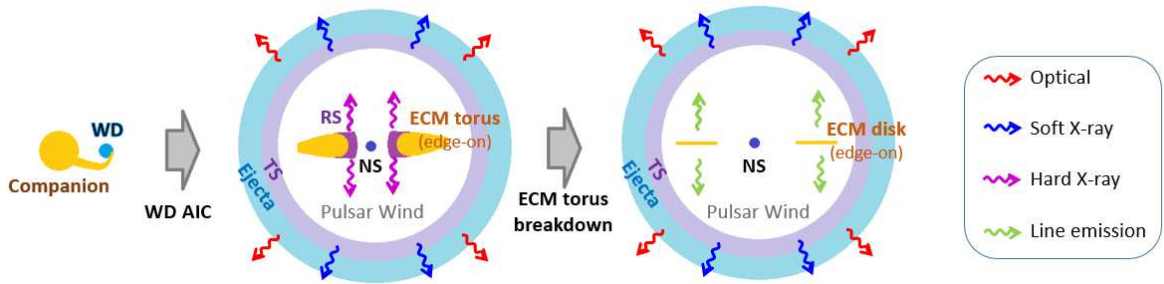}
\caption{Illustration of a post-AIC system (not to scale), including the stages before and after the breakdown of an ECM torus.}\label{illus}
\end{center}
\end{figure*}

In principle, the AIC NSs could sometimes (but may not always) be converted into magnetars, as these NSs can initially rotate very rapidly and differentially. Following this consideration, it has been suggested that the rapid spin-down of the magnetars can substantially affect the AIC optical transient, e.g., by providing an extra energy source to heat the AIC ejecta (Yu et al. 2015, 2019). In more detail, the energy released from the magnetars can be initially in the form of a Poynting flux, which could eventually transform into a relativistic electron-positron wind. Then, questions arise as to what kind of emission can be produced by the pulsar wind (PW) and how
energy is transferred from the PW to the AIC ejecta. It is expected that the PW emission, if detected, can help to identify AIC candidates in the future.

Basically, PW emission could be mainly determined by the collisions of the PW with the AIC ejecta and, in particular, with the companion star. An illustration of these interactions is presented in Figure \ref{illus}. Considering the fact the spin-down luminosity of a magnetar can be as high as $\sim 10^{45}\rm erg~s^{-1}$, the companion could be completely evaporated by the PW in a few days, because the gravitational binding energy of the
companion is probably not much higher than $\sim10^{48}$ erg. The evaporated companion material (ECM) would finally form a torus due to the orbital motion, which surrounds the magnetar and blocks the PW in the orbital plane. The formation of such an ECM torus and its interaction with the PW are investigated in Section 2.
Meanwhile, the termination shock (TS) of the PW due to its interaction with the AIC ejecta is calculated in Section 3. Both of these two PW emission components can penetrate the AIC ejecta and become detectable as long as the ejecta becomes transparent for X-rays. A summary of our calculations is given in Section 4. In Section 5, we discuss the possible application of the AIC model in explaining the multi-wavelength emission of a recently discovered fast-evolving luminous optical transient: AT 2018cow.

\section{The interaction of a PW with a companion}
\subsection{The evaporation of a companion}
Some NSs formed from AIC events are considered to rotate very rapidly and possibly differentially. In this case, the magnetic fields of the NSs can be significantly amplified through magnetorotational instability (Duncan \& Thompson 1992; Dessart et al. 2007; Cheng \& Yu 2014). However, in practice, it is still difficult to assign a fiducial value to the spin periods and the magnetic fields of the NSs, because these parameter values are highly dependent on the properties of progenitor WDs. In principle, the surface magnetic field of an AIC NS could become as high as $\sim10^{15}$ G, if the progenitor
WD is rotating extremely and is highly super-Chandrasekhar (Dessart et al. 2007). However, the rate of such extreme AICs is likely to be very low, because of the strong constraints from their yields of \textit{r}-process nuclei. Therefore, for a more general consideration, here we adopt moderate values of $B_{\rm p}\sim3\times10^{14}$ G and $P_{\rm i}\sim5$ ms for the dipolar magnetic field strength and the initial spin period of the NS, respectively. An additional motivation of this parameter choice is to make the AIC events relevant to the observed fast-evolving luminous transients (e.g., Yu et al. 2015), with the following spin-down luminosity and timescale:
\begin{eqnarray}
L_{\rm sd,i}=1.4\times10^{45}~B_{\rm
p,14.5}^{2}P_{\rm i,-2.3}^{-4}\rm ~erg~s^{-1}
\end{eqnarray}
and
\begin{eqnarray}
t_{\rm sd}=6.4~B_{\rm p,14.5}^{-2}P_{\rm i,-2.3}^{2}\rm
~day,\end{eqnarray}
where the conventional notation $Q_{x}=Q/10^x$ is used in cgs units. As usual, the NS spin-down is simply considered to be dominated by magnetic dipole radiation and thus the temporal dependence of the spin-down luminosity can be written as $L_{\rm sd}(t)=L_{\rm sd,i}(1+{t/ t_{\rm sd}})^{-2}$.

Due to the high pressure of the PW acting on the companion star, the star can be evaporated at a rate of (van den
Heuvel \& van Paradijs 1988)
\begin{eqnarray}
\dot{M}_{\rm ev}=f\left(R_{\rm \star}\over a_{}\right)^2{L_{\rm sd}\over 2v_{\rm esc}^2},
\end{eqnarray}
where $f$ represents the faction of the wind energy that can be transferred to the companion, the value of which is determined by the dynamical interaction between the wind and the companion. When an AIC happens, the companion star must have filled the Roche lobe. Thus, we can use the radius of the Roche lobe to represent the radius of the companion; i.e., the companion radius can be connected with the orbital radius by $R_{\star}/a=0.38+0.2\log_{10}q$, where $q$ is the mass ratio of the companion to the NS (Eggleton 1983). For an escaping velocity of the companion as
\begin{eqnarray}
v_{\rm esc}=\sqrt{2GM_{\star}\over R_{\star}}=280\left({M_{\star}\over M_{\odot}}\right)^{1/2}\left({R_{\star}\over 5R_{\odot}}\right)^{-1/2}~{\rm km~s^{-1}},
\end{eqnarray}
the timescale on which the companion can be totally disrupted by the PW can be estimated to (for $q=1$)
\begin{eqnarray}
t_{\rm ev}={M_{\star}\over \dot{M}_{\rm ev}}=2.4~f_{-1}^{-1}\left({M_{\star}\over M_{\odot}}\right)^2\left({R_{\star}\over 5R_{\odot}}\right)^{-1}L_{\rm sd,45}^{-1}{\rm~ days},
\end{eqnarray}
where $M_{\star}$ is the mass of the companion. Accompanying with the companion evaporation, the ECM gradually constitutes a torus through its orbital motion with a period of (for $q=1$)
\begin{eqnarray}
t_{\rm orb}=2\pi \sqrt{a_{}^3\over GM_{\rm ns}}=4.7~\left({R_{\star}\over 5R_{\odot}}\right)^{3/2}{\rm ~days},
\end{eqnarray}
where the mass of the NS is taken as $M_{\rm ns}=1.4M_{\odot}$. Roughly, this ECM torus can have a height of $h\sim 2R_{\star}$ and a width of $\Delta\sim v_{\rm esc}t_{\rm ev}=5.8\times10^{12}~f_{-1}^{-1}\left({M_{\star}/ M_{\odot}}\right)^{5/2}\left({R_{\star}/ 5R_{\odot}}\right)^{-3/2}L_{\rm sd,45}^{-1}~{\rm cm}$.

Because of the appearance of the ECM torus, the PW that blew from the central NS would be blocked in the orbital plane of an opening angle of $\sim h/r_{\rm in}$, where $r_{\rm in}$ is the inner radius of the torus. The collision between the PW and the torus can drive a pair of shocks, including a forward shock (FS) propagating into the ECM and a reverse shock (RS) propagating into the PW. Due to the high mass and high density of the ECM torus, the velocity of the FS relative to the unshocked torus cannot be much higher than the torus velocity, i.e., the escaping velocity. The timescale of the FS breaking out from the torus can be estimated by $t_{\rm bo}\sim2M_{\star}v_{\rm esc}^2/(f L_{\rm sd})=0.4f_{\rm -1}^{-1}(M_{\star}/M_{\odot})^2L_{\rm sd, 45}^{-1}(R_{\star}/5R_{\odot})^{-1}$ day (Li \& Yu 2016), which is shorter than $t_{\rm ev}$ and $t_{\rm orb}$.
Therefore, the energy accumulated by the FS is very limited. What is of more interest is the emission property of the longer-lasting RS.

\subsection{The dynamics of a RS}

After the FS crosses the ECM torus, the dynamical evolution of the RS is still highly related to the evolution of the torus, as the shocked wind can energize the torus by irradiating it and pushing it outward. The fraction of the RS emission absorbed by the torus can be estimated by $\sim h/2r_{\rm in}$, while the remaining large fraction of the emission can freely escape from the RS region through high latitudes. Therefore, with the rapid increase of $r_{\rm in}$, the radiation heating effect can in fact be ignored for the torus. The energy transfer from the shocked wind to the torus is achieved mainly through their mechanical contact. On the one hand, the high pressure of the shocked wind acting on the inner boundary of the ECM torus can heat the torus by compressing it. On the other hand, the increasing internal energy of the torus means that the torus tends to
expand more quickly and even freely. These compression and
expansion tendencies can finally achieve a balance, as the torus
is actually embedded in the flow-field environment of the PW. In this case, the bulk kinetic energy and internal energy of the ECM torus are considered to satisfy an equipartition. Then the total energy of the torus can be written as $E_{\rm ecm}=M_{\star}v_{\rm ecm}^2$, where
\begin{eqnarray}
v_{\rm ecm}={dr_{\rm in}\over dt}\label{vecm}
\end{eqnarray}
is used as an uniform velocity.

The dynamical evolution of the ECM torus can be determined by the work done by the PW as ${dE_{\rm ecm}/dt}=2\pi r_{\rm in}hP_{\rm rs}v_{\rm ecm}$, where $P_{\rm rs}$ is the pressure of the RS. Then we can have
\begin{eqnarray}
{dv_{\rm ecm}\over dt}={\pi r_{\rm in}hP_{\rm rs}\over M_{\star}}.\label{Ekt2}
\end{eqnarray}
Here, for simplicity, it is assumed that the torus existed before
the shock interaction. This leads to somewhat of an overestimation
of the work in the early days. During this period, the
shock interaction actually takes place along with the companion
evaporation and the torus circularization. According to the
jump conditions of a relativistic shock, the shocked wind
pressure can be derived to
\begin{eqnarray}
P_{\rm rs}&=&{1\over 3}e_{\rm rs}={4\over 3}{\Gamma'}^2_{\rm rs}{n'}_{\rm e}m_{\rm e}c^2={L_{\rm sd}\over 3\pi r_{\rm rs}^2c},
\end{eqnarray}
where $e_{\rm rs}$ is the energy density of the shocked wind, ${\Gamma'}_{\rm rs}$ is the shock Lorentz factor measured in the comoving frame of the unshocked wind, and $n'_{\rm e}=L_{\rm sd}/(4\pi r_{\rm rs}^2\Gamma_{\rm w}^2m_{\rm e}c^3)$ is the comoving electron number density in front of the RS. In this Letter, we always have ${\Gamma'}_{\rm rs}\approx \Gamma_{\rm w}$, where $\Gamma_{\rm w}$ is the bulk Lorentz factor of the unshocked wind. The value of $\Gamma_{\rm w}$ is difficult to determine in theory, but could be inferred from some observations (e.g., the Crab pulsar; Atoyan 1999) to be about $\sim10^4-10^7$.  Finally, the radius of the RS, $r_{\rm rs}$, can be determined by according to the mechanical balance between the shocked wind and the ECM torus; i.e., we can have
\begin{eqnarray}
r_{\rm rs}=\left({L_{\rm sd}\over \pi \rho_{\rm ecm} v_{\rm ecm}^2c}\right)^{1/2}\label{rrs}
\end{eqnarray}
from the equation $P_{\rm rs}=P_{\rm ecm}={1\over3}\rho_{\rm ecm}v_{\rm ecm}^2$, where the density of the torus can be given by
\begin{eqnarray}
\rho_{\rm ecm}={M_{\star}\over \pi[(r_{\rm in}+\Delta)^2-r_{\rm in}^2]h}.\label{rhoecm}
\end{eqnarray}

\begin{figure}
\begin{center}
\includegraphics[scale=0.4]{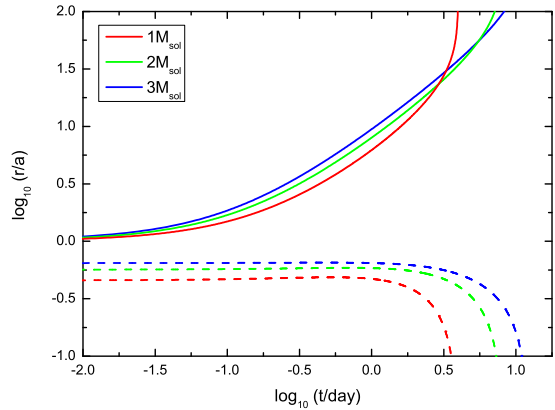}
\caption{Evolution of the inner radius of the ECM torus (solid lines) and the radius of the RS front (dashed lines), for three typical companion masses as labeled. The other parameters are taken as $R_{\star}=5R_{\odot}$, $L_{\rm sd,i}=10^{45}\rm erg~s^{-1}$, $t_{\rm sd}=5$ day, and $\Gamma_{\rm w}=3\times10^4$.
}\label{rfr}
\end{center}
\end{figure}

By solving Equations (\ref{vecm}$-$\ref{rhoecm}) together, we plot the evolution of the radii of the ECM torus and the RS in Figure \ref{rfr}, for three typical masses of the companion star. While the torus moves outward at a speed of $\sim 10^3\rm km~s^{-1}$, the RS is nearly standing for a period of several to 10 days. Nevertheless,
with the continuous accumulation of the shocked wind
material, the RS front eventually moves inward very quickly.
As a result, the wind pressure is finally increased enough so
that it breaks through the blockage of the wind by the torus. It
is possible that, at this moment, the inner boundary of the ECM
torus is destroyed completely and all of the shocked wind
material pours out through the upper and lower edges, although
these processes cannot be described in detail in our simple
model. In any case, the breakdown time of the torus can be
found to depend on the torus mass, i.e., the companion mass.
The more massive the companion, the longer-lasting the RS
and its emission. After the torus breakdown, the interface
between the PW and the torus is streamlined smoothly, the
torus becomes an extremely thin ring-like disk, and the RS
disappears.

\begin{figure}
\begin{center}
\includegraphics[scale=0.4]{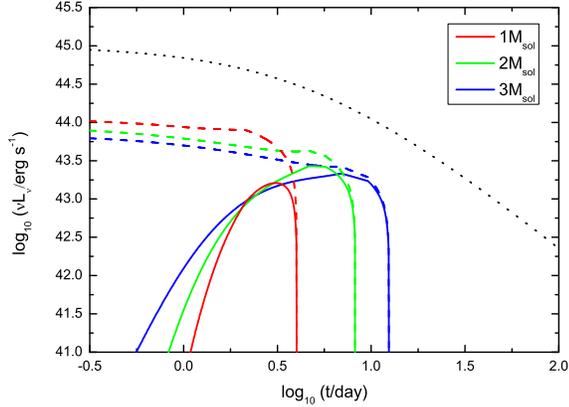}
\caption{The RS synchrotron emission at a hard X-ray frequency of $h\nu=50$ keV, corresponding to the dynamical evolutions presented in Figure \ref{rfr} and microphysical parameters $\epsilon_{\rm B}=0.01$ and $p=2.3$. The dashed and solid lines are obtained without and with the ejecta absorption effect, respectively, for ejecta parameters $M_{\rm ej}=0.005M_{\odot}$, $v_{\rm ej}=0.1c$, and $\kappa_{\nu}({\rm 50keV})=0.36\rm cm^2g^{-1}$. The dotted line represents the input spin-down luminosity.
}\label{RSemission}
\end{center}
\end{figure}

\subsection{The synchrotron emission of an RS}
The number of the electrons and positrons accumulated by the RS increases at a rate of
\begin{eqnarray}
{dN_{\rm e}\over dt}&=&2\pi r_{\rm rs}h (v_{\rm w}-v_{\rm rs}){\Gamma'}_{\rm rs}n'_{\rm e}\nonumber\\
&\approx& {h\over 2 r_{\rm rs}}{L_{\rm sd}\over\Gamma_{\rm w}m_{\rm e}c^2},
\end{eqnarray}
where $v_{\rm w}\approx c$ and $v_{\rm rs}\ll c$ are the velocities of the unshocked wind and the RS, respectively. The second expression above can be easily understood as a fraction of $h/2r_{\rm rs}$ of the total particle flux $\dot{N}_{\rm e}=L_{\rm sd}/\Gamma_{\rm w}m_{\rm e}c^2$ of the PW. These electrons can initially all be
accelerated by the RS and distributed with their random Lorentz factors as $dN_{\rm e}/d\gamma\propto \gamma^{-p}$ for $\gamma\geqslant\gamma_{\rm m}=[(p-2)/(p-1)]\Gamma'_{\rm rs}$, where $p$ is a constant spectral index. However, immediately the overwhelming majority of these electrons will cool down to be non-relativistic (i.e., $\gamma_{\rm }\lesssim 2$) because of their synchrotron radiation. By defining a cooling timescale for relativistic electrons as $t_{\rm col}=3\pi m_{\rm e}c/\sigma_{\rm T}B^2$, the remaining number of relativistic electrons at the time $t$ can be estimated by $N_{\rm e,rel}=N_{\rm e}t_{\rm col}/t$, where $\sigma_{\rm T}$ is the Thomson cross section, $B=(4\pi\epsilon_{\rm B}e_{\rm rs})^{1/2}$ is the strength of the stochastic magnetic field for an equipartition factor $\epsilon_{\rm B}$, and $N_{\rm e}\approx \dot{N}_{\rm e}t$ is the total number of radiating electrons at that time.

Following the electron distribution, the luminosity at a frequency $\nu$ of the synchrotron emission of the RS can be given by (Sari et al. 1998)
\begin{equation}
L_{\nu}^{}=L_{\nu,\max}^{}\times\left\{
\begin{array}{ll}
\left({\nu\over\nu_{\rm p}}\right)^{-1/2},&\nu_{\rm p}<\nu<\nu_{\rm m};\\
\left({\nu_{\rm m}\over\nu_{\rm p}}\right)^{-1/2}\left({\nu\over\nu_{\rm m}}\right)^{-p/2},&\nu>\nu_{\rm m}.\\
\end{array}\right.\label{Lsyn}
\end{equation}
Here the peak spectral power is given by $L_{\nu,
\max}^{}= N_{\rm e,rel}{m_{\rm e}c^2\sigma_{\rm T}B/ 3q_{\rm e}}$ and the two characteristic frequencies are defined as $\nu_{\rm p}={2q_{\rm e}B/\pi m_{\rm e}c}$ corresponding to $\gamma=2$ and $\nu_{\rm m}={q_{\rm e}B\gamma_{\rm m}^2/2\pi m_{\rm e}c}$, where $q_{\rm e}$ is the electron charge. Obviously, the peak of the $\nu L_{\nu}$ spectrum of the RS emission is at the frequency $\nu_{\rm m}$, which falls in the hard X-ray band for typical model parameters as (for $t<t_{\rm sd}$):
\begin{equation}
\nu_{\rm m}=1.0\times10^{19}L_{\rm sd,45}^{1/2}\Gamma_{\rm w,4.5}^2\epsilon_{\rm B, -2}^{1/2}r_{\rm rs, 12}^{-1}~{\rm Hz}.
\end{equation}
The corresponding peak luminosity can be estimated to
\begin{equation}
[\nu_{\rm }L_{\nu_{\rm }}]_{\rm m}=8.0\times10^{43}L_{\rm sd,45}\left({R_{\star}\over 5R_{\odot}}\right)r_{\rm rs,12}^{-1}~\rm erg~s^{-1}.
\end{equation}
In principle, this synchrotron emission could be boosted further
by inverse-Compton scattering to much higher energies, typically as $2\gamma_{\rm m}^2h\nu_{\rm m}\sim 4L_{\rm sd,45}^{1/2}\Gamma_{\rm w,4.5}^4\epsilon_{\rm B, -2}^{1/2}r_{\rm rs, 12}^{-1}$ TeV. This photo nenergy, which is much higher than $\gamma_{\rm m}m_{\rm e}c^2\sim3.5$ GeV indicates that the inverse-Compton scattering is actually
seriously suppressed by the Klein¨CNishina effect.

It should be noted that the RS emission probably can be
absorbed by the surrounding AIC ejecta, as long as the ejecta
are opaque. Therefore, a leakage coefficient needs to be invoked to calculate the observable luminosity as
\begin{equation}
L_{\nu}^{\rm obs}=L_{\nu}{\rm e}^{-\tau_{\nu}}, \label{Lhx}
\end{equation}
where $\tau_{\nu}=3\kappa_{\nu}M_{\rm ej}/4\pi r_{\rm ej}^2$ is the optical depth of the ejecta, which is highly frequency-dependent, and $M_{\rm ej}$ and $r_{\rm ej}$ are the mass and the outer radius of the AIC ejecta, respectively. Generally, the opacity $\kappa_{\nu}$ for the X-rays from $\sim$100 eV to a few tens of keV is dominated by photoelectric absorption, while very hard X-rays experience predominantly Compton
scattering. The specific opacity values are sensitive to the composition and ionization degree of the ejecta. In our calculations, we adopt the following empirical expression as (for $h\nu\gtrsim$ 0.1keV):
\begin{eqnarray}
\kappa_{\nu}&=&C_1\left({h\nu\over {\rm 1 keV}}\right)^{-3}+C_2\left[\left({h\nu\over {\rm 50keV}}\right)^{-0.36}+\left({h\nu\over {\rm 50keV}}\right)^{0.65}\right]^{-1}
\nonumber\\
&&+C_3
\end{eqnarray}
with the coefficients $C_1=2.0{\rm cm^2g^{-1}}$, $C_2=0.7{\rm cm^2g^{-1}}$, and $C_3=0.012{\rm cm^2g^{-1}}$. This expression is obtained by fitting the numerical results in Figure 8 of Kotera et al. (2013). Then, in Figure \ref{RSemission} we plot the hard X-ray light curves of the RS emission given by Equations (\ref{Lsyn}) and (\ref{Lhx}) for different companion masses, where $\kappa_{\nu}=0.36\rm cm^2g^{-1}$ is adopted for the selected photon energy of $h\nu=50\rm keV$. The higher the mass, the longer the RS emission, as the final abrupt decay of this emission is due to the breakdown of the ECM torus.

Throughout this Letter, the AIC ejecta is simply assumed to be isotropic. However, in reality, the ejecta is probably anisotropic and even nonexistent at some directions (e.g., the low-latitude areas). Therefore, in principle, the RS emission can freely leak from these directions and be detected at very early times.

\begin{figure}
\begin{center}
\includegraphics[scale=0.4]{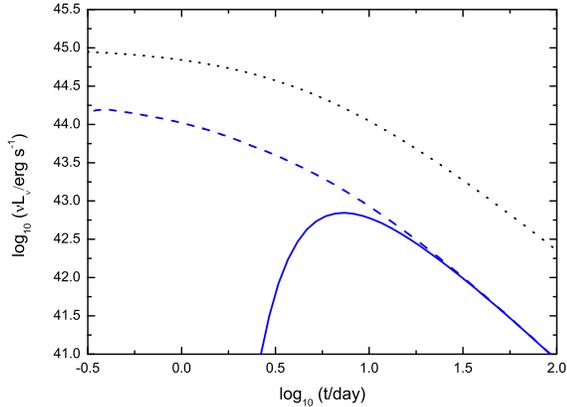}
\caption{TS synchrotron emission at a soft X-ray frequency of $h\nu=1$ keV. The dashed and solid lines are obtained without and with the ejecta absorption effect, respectively. The ejecta parameters are taken as $M_{\rm ej}=0.005M_{\odot}$, $v_{\rm ej}=0.1c$, and $\kappa_{\nu}(1{\rm keV})=2.2\rm cm^2g^{-1}$. The dotted line represents the input spin-down luminosity.}\label{TSemission}
\end{center}
\end{figure}

\section{The emission of a TS and a hot ejecta}
While the PW in the orbital plane is blocked by the ECM torus, the wind at high latitudes can simultaneously collide with the preceding AIC ejecta to form a TS. In comparison with the RS emission discussed above, the TS emission can have a much longer duration and fall into relatively lower energy bands because the radius of the TS is much larger than that of the RS.

As in the previous section, in order to calculate the TS
emission we need to determine the dynamical evolution of the AIC ejecta simultaneously. Different from the ECM torus, the AIC ejecta can experience a free expansion, as long as $M_{\rm sw}\ll M_{\rm ej}$
where $M_{\rm sw}$ is the mass of the swept-up circumstellar medium (CSM). Therefore, the internal energy of the ejecta $E_{\rm int}$ is radiation dominated and its evolution is determined by
\begin{eqnarray}
\frac{ d E_{\rm int}}{d t} =  - L_{\rm th}- 4\pi r_{\rm ej}^2P_{\rm ej}v_{\rm ej}+L_{\rm h} .\label{Eint1}
\end{eqnarray}
The meanings of the three terms on the right side of the equation are explained as follows. Firstly, the bolometric luminosity $L_{\rm th}$ of the ejecta thermal emission can be estimated by (Kasen \& Bildsten 2010; Yu \& Li 2017)
\begin{equation}
L_{\rm th}={E_{\rm int}c\over  r_{\rm ej}\tau_{\rm }}\left(1-{\rm e}^{-\tau_{\rm }}\right),\label{Lbol}
\end{equation}
where the optical depth $\tau_{\rm }$ is given for an electron-scattering opacity $\kappa_{\rm es}=0.2\rm ~cm^2~g^{-1}$ as the thermal emission is mainly in the ultraviolet (UV)/optical bands (see Equation \ref{temperature}). Second, the term $4\pi r_{\rm ej}^2P_{\rm ej}v_{\rm ej}$
represents the adiabatic cooling of the ejecta with $P_{\rm ej}$ and $v_{\rm ej}$ being the pressure and velocity of the ejecta, respectively. Due to this adiabatic expansion, the velocity of the ejecta can be somewhat increased. However,
this acceleration effect is insignificant, because the initial velocity is high. Therefore, in our calculations we take a constant $v_{\rm ej}$ and use $r_{\rm ej}=v_{\rm ej}t$ directly. Finally, the heating power $L_{\rm h}$ in Equation (\ref{Eint1}) is mainly related to the absorption of the PW emission by the ejecta.

For given $E_{\rm int}$ and $r_{\rm ej}$, the energy density of the TS region can be obtained by
\begin{eqnarray}
e_{\rm ts}= {e}_{\rm ej}={3E_{\rm int}\over 4\pi r_{\rm ej}^3}.
\end{eqnarray}
Then, similar to the treatments for the RS in the previous section, the synchrotron emission of the TS can be calculated, which peaks around
\begin{equation}
\nu_{\rm m}=1.0\times10^{17}L_{\rm sd,45}^{1/2}\Gamma_{\rm w,4.5}^2\epsilon_{\rm B, -2}^{1/2}r_{\rm rs, 14}^{-1}~{\rm Hz},
\end{equation}
for the first few days. Subsequently, with the expansion of the TS region, the cooling timescale of the relativistic electrons in the TS region can finally become longer than the dynamical time. Then a cooling frequency should be extra introduced to fix the synchrotron spectra, i.e., $\nu_{\rm c}={q_{\rm e}B\gamma_{\rm c}^2/(2\pi m_{\rm e}c)}$ with $\gamma_{\rm c}=6\pi m_{\rm e}c/(\sigma_{\rm T}B^2t)$ (see Sari et al. 1998 for details). In addition, with the softening of
the synchrotron photons, the Klein¨CNishina suppression on their
inverse-Compton scattering can be gradually relieved; in particular, if the Lorentz factor of the PW is obviously smaller than $\sim10^4$. In this Letter, we do not take this Low-$\Gamma_{\rm w}$ situation into account. In any case, as the luminosity $L_{\nu}$ of the TS emission is obtained, the heating power $L_{\rm h}$ of the ejecta can be subsequently derived from the following integral:
\begin{eqnarray}
L_{\rm h}&=&\int\left(1-{\rm e}^{-\tau_{\nu}}\right)L_{\nu}d\nu.
\end{eqnarray}
In principle, the heating power can also be contributed by the radioactive decay of nickel. However, this radioactive power is actually drastically smaller than the PW luminosity and can be ignored, because of the very limited ejecta mass.

By using the above equations, we can simultaneously calculate the nonthermal emission of the TS, $L_{\nu}$, and the thermal emission of the AIC ejecta, $L_{\rm th}$. On the one hand, in Figure \ref{TSemission} we provide an example soft X-ray light curve of the TS emission, where $\kappa_{\rm \nu}=2.2\rm ~cm^2g^{-1}$ is taken for the selected photon energy of $h\nu\sim1$ keV. On the other hand, for the ejecta thermal emission, here we only present an effective temperature as
\begin{eqnarray}
T_{\rm p}&=&\left({L_{\rm th,p}\over 4\pi v_{\rm ej}^2t_{\rm p}^2}\right)^{1/4}\nonumber\\
&&=6.2\times10^{4}\left({L_{\rm th,p}\over 5\times10^{44}{\rm erg~s^{-1}}}\right)^{1/4}\nonumber\\
&&\times\left({M_{\rm ej}\over 5\times10^{-3}M_{\odot}}\right)^{-1/4}\left({v_{\rm ej}\over 0.1c}\right)^{-1/4} \rm K,\label{temperature}
\end{eqnarray}
for a general impression, because this emission component has been investigated in detail in previous works (e.g., Yu et al. 2015, 2019). This temperature is obtained by assuming a blackbody spectrum and with a peak emission time of $t_{\rm p}=(3\kappa_{\rm es} M_{\rm ej}/4\pi c v_{\rm ej})^{1/2}$. The ejecta thermal emission is mainly in the UV to optical bands.

\section{Summary}
\begin{figure}
\begin{center}
\includegraphics[scale=0.4]{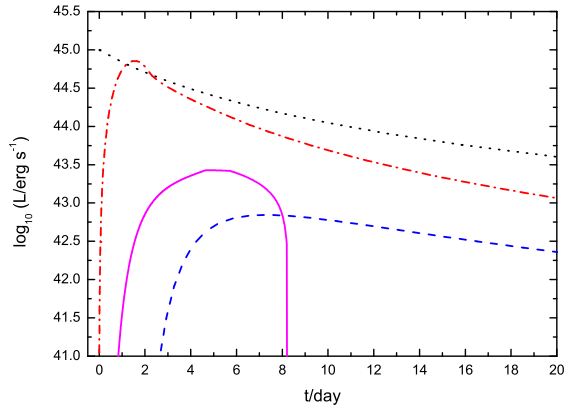}
\caption{Light curves of the AIC transient emissions, including the ejecta thermal emission (bolometric; dashed-dotted), the RS hard X-ray emission (50 keV; solid), and the TS soft X-ray emission (1 keV; dashed). The ejecta absorption effect is taken into account. The dotted line represents the input spin-down luminosity. A companion mass of $M_{\star}=2M_{\odot}$ is taken for the RS emission and the other parameters are same as those used in the previous figures.}\label{LCmul}
\end{center}
\end{figure}

\begin{figure}
\begin{center}
\includegraphics[scale=0.4]{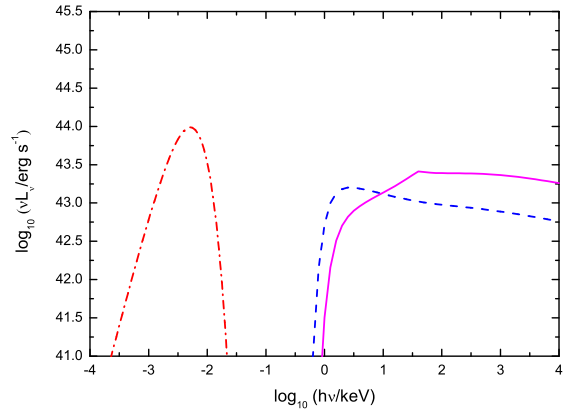}
\caption{Spectra of the AIC transient emissions for day 5 in Figure \ref{LCmul}. The dashed¨Cdotted, dashed, and solid lines correspond to the ejecta blackbody emission, the TS synchrotron emission, and the RS synchrotron emission, respectively.  The ejecta absorption effect is taken into account, which drastically suppresses the soft X-ray emission.}\label{SP}
\end{center}
\end{figure}

The AICs of accreting WDs can launch a low-mass ejecta and sometimes give birth to a millisecond magnetar. The interactions of a magnetar-driven PW with the AIC ejecta and the companion star can lead to various multi-wavelength transients, which can be regarded as the observational signatures of the AIC events.
First of all, because of the heating by the PW emission and as well as by radioactivity, the hot AIC ejecta can give rise to a fast-evolving and luminous UV/optical transient, which is the most conspicuous signature. In this
Letter, we have found that hard and soft X-ray emissions
accompanying this UV/optical transient can also be produced
by the PW itself, which is powered by an RS and a TS. Example light curves of these multi-wavelength AIC transients are presented in Figure \ref{LCmul}, while their spectra at day 5 are presented in Figure \ref{SP}. As a result of the ejecta absorption, the soft X-ray emissions of both the RS and the TS are suppressed substantially. It is worth emphasizing that, different from the long-lasting emissions of the TS and the hot ejecta, the hard X-ray emission from the RS only appears in the first few weeks
before it decays abruptly due to a torus breakdown. As a further
consideration, after the torus breakdown the ECM still exists in
a very thin and ring-like disk. This disk can be irradiated by the
soft X-rays from the TS and thus some hydrogen and helium
lines can be expected. These lines, occurring after a hard X-ray
decay, provide a crucial signature for the AIC events. In
principle, at very late times the TS could extra contribute a
high-energy (e.g., $\sim$GeV) gamma-ray emission if the synchrotron
self-Compton scattering takes place successfully and
effectively.

Additionally, the hard X-ray emission arising from the companion-interaction can make the AIC events different from the analogous phenomena such as NS mergers (Yu et al. 2013; Metzger \& Piro 2014) and WD mergers (Yu et al. 2019). It is
suggested that these merger events are also able to create a system consisting of a millisecond magnetar and a low-mass ejecta.

\section{Discussion}

It is interesting to associate the AIC model with recently discovered fast-evolving luminous transients (e.g., Yu et al. 2015) and, in particular, with the unusual transient AT 2018cow. This transient was observed in all of the optical, soft, and hard X-ray bands (Prentice et al. 2018; Rivera Sandoval et al. 2018; Perley et al. 2019; Kuin et al. 2019; Margutti et al. 2019). The combination of these different emission components roughly tracks a temporal behavior around $t^{-2}$, which is well consistent with the spinning-down magnetar (Fang et al. 2018; Prentice et al. 2018; Margutti et al. 2019). In more detail, the hard X-ray emission of AT 2018cow was only detected during the first 10 days and disappeared abruptly, which is in good agreement with the expectation of the RS emission in our model. Coincidentally, hydrogen and helium lines were
identified in the emission after the disappearance of the hard
X-rays. The velocity inferred from these lines also agrees with
the origin of the remaining ECM disk. Finally, it was
discovered that radio emission from AT 2018cow increased continuously during the first hundred days and reached a turnover later (Ho et al. 2019). In the AIC model, this radio emission can be naturally explained by the shock interaction between the AIC ejecta and the CSM, where the medium had been substantially shaped by the stellar wind of the companion before the AIC event (Piro \& Kulkarni 2013; Moriya 2016). In principle, this CSM interaction could also partially contribute to the optical transient emission, as suggested by Fox \& Smith (2019).

In any case, for accurate fittings to the observational data of AT 2018cow, our model still needs to be modified and improved, e.g., by invoking an anisotropic structure and a radial distribution of the ejecta. It would also be best to calculate the shock dynamics by simultaneously incorporating the companion evaporation and torus circularization. These model improvements can help to distinguish our model from other analogous models for AT 2018cow, i.e., a WD tidal disruption event (Kuin et al. 2019),  a jet driven by an accreting NS colliding with a giant star (Soker et al.
2019), and, in particular, an electron-capture collapse of a remnant of a WD merger (Lyutikov \& Toonen
2018).

\acknowledgements
The authors thank Zi-Gao Dai and Bing Zhang for their useful comments.
This work is supported by the National Natural Science
Foundation of China (grant Nos. 11822302, 11833003, 11773015, and U1838201),
and the National Key Research and Development Program of China
(2016YFA0400803).

\end{document}